\documentclass[runningheads]{llncs}
\usepackage{amsfonts, enumerate}
\usepackage{graphicx, amsmath, amssymb}

\newcommand{\cO}{\mathcal{O}}

\newcommand{\const}{0.38685}
\newcommand{\nrSep}{1.6181}
\newcommand{\nrPmc}{1.7549}
\newcommand{\boundPol}{2.6151}
\newcommand{\cC}{\mathcal{C}}
\newcommand{\constB}{0.2263}

\newcommand{\sm}{\setminus}
\newcommand{\mB}{{\mathcal{B}}}

\begin{document}
\title{Treewidth computation and extremal combinatorics\thanks{This research was partially
 supported by the Research Council of Norway.}}

\author{Fedor V. Fomin \and  Yngve Villanger}

\begin{small}
\institute{Department of Informatics, University of Bergen,\\
 N-5020 Bergen,Norway \{fedor.fomin,yngve.villanger\}@ii.uib.no}
 \end{small}
\maketitle

\begin{abstract}
For a given graph $G$ and integers $b,f \geq 0$,  let $S$ be a subset of vertices of $G$ of size $b+1$
such that the subgraph of $G$ induced by $S$ is connected and $S$ can be separated from other vertices of $G$ by removing
 $f$ vertices. We prove that every graph on $n$ vertices contains at most  $n\binom{b+f}{b}$ such vertex subsets.
This result from extremal
combinatorics appears to be very useful in the design of several enumeration and exact algorithms.  In particular, we use it to
provide algorithms that for a given $n$-vertex graph $G$
\begin{itemize}
\item compute the treewidth of $G$ in time $\cO(\nrPmc^n)$ by making use of  exponential space and
in time $\cO(\boundPol^n)$ and polynomial space;
\item decide  in time $\cO(({\frac{2n+k+1}{3})^{k+1}\cdot kn^6})$ if the treewidth of $G$  is at most $k$;
\item list all minimal separators of $G$  in time $\cO(\nrSep^n)$ and
 all potential maximal cliques of $G$ in time $\cO(\nrPmc^n)$.
\end{itemize}
This significantly improves previous 	algorithms  for these
problems.
\end{abstract}

\section{Introduction}
The aim of \emph{exact algorithms} is to optimally solve hard
problems exponentially faster than brute-force search.
 The first papers in the area date back to the sixties and seventies \cite{HeldKarp62jsiam,TarjanTrojanowski77}.
 For the last two decades the  amount of literature
 devoted to this topic has been tremendous and
it is impossible to give here a  list of representative references without missing significant results.
Recent surveys  \cite{FominGrandoniKratsch05beatcs,Iwama04beatcs,Schoening05stacs,Woeginger03} provide a comprehensive
information on exact algorithms.
It is very natural to assume the existence of strong links between
the area of exact algorithms and some areas of extremal
combinatorics, especially the part of extremal combinatorics which
studies the maximum (minimum) cardinalities of a system of subsets
of some set satisfying certain properties. Strangely enough, there
are not so many examples of such links in the literature, and the majority of
exact algorithms are based on the so-called branching
(backtracking) technique which traces back to the works  of Davis,
Putnam, Logemann, and  Loveland  \cite{DavisLL62,DavisP60-A}.

In this paper,  we prove a  combinatorial lemma which appears to
be very useful in the analysis   of certain enumeration and exact
 algorithms. For a vertex $v$ of a graph $G$ and
integers $b,f\geq 0$, let $t(b,f)$ be the maximum  number of
connected induced subgraphs of $G$ of size $b+1$ such that the
intersection of all these subgraphs is nonempty  and each such a
subgraph has exactly $f$ neighbors (a neighbor of a subgraph $H$
is a vertex of $G\setminus H$ which is adjacent to a vertex of
$H$). Then the combinatorial lemma states that  $t(b,f)\leq
\binom{b+f}{b}$ (and it is easy to check that this bound is
tight).
 This  can be seen as a variation
of Bollob{\'a}ss Theorem \cite{Bollobas65}, which is one of the
corner-stones in extremal set theory. (See Section 9.2.2 of
\cite{Jukna01Ex} for detailed discussions on  Bollob{\'a}ss
Theorem and its variants.)

We use this combinatorial result to obtain faster algorithm
for a number of problems related to the treewidth of a graph.
The treewidth is a fundamental graph parameter from Graph Minors
Theory by Robertson and Seymour \cite{RobertsonS86} and it has
numerous algorithmic applications, see  the  surveys
\cite{Bodlaender98,Bodlaender08}.
The problems to compute the treewidth is known to
be NP-hard  \cite{ArnborgCP87} and the best known approximation
algorithm for treewidth has  a factor $ \sqrt{\log OPT}$
\cite{FeigeHL05}.
It is an old open question whether the treewidth
can be approximated within a constant factor.
Treewidth is known
to be fixed parameter tractable. Moreover, for any fixed $k$,
there is a linear time algorithm due to Bodlaender  \cite{Bodlaender96}  computing the treewidth of
graphs of treewidth at most $k$. Unfortunately, huge
hidden constants in the running time of Bodlaender's
algorithm is a serious obstacle to its implementation.
For small values of $k$, the classical algorithm of  Arnborg, Corneil and Proskurowski
\cite{ArnborgCP87} from 1987 which runs in time $\cO(n^{k+2})$  can be used to decide if the treewidth of a graph is at most $k$. 
The first exact algorithm computing the treewidth of an $n$-vertex
graph is due to Fomin et al. \cite{FominKT04} and has running time
$\cO({1.9601}^n)$. Later these results were improved in
\cite{FominKratschTodincaV08,Villanger06} to $\cO({1.8899}^n)$.
 Both algorithms
use exponential space. The fastest polynomial space algorithm for treewidth
prior to this work  is due to Bodlaender et al. \cite{BodlaenderFKKT06} and runs in time $\cO(2.9512^n)$.

\medskip\noindent\textbf{Our results.}
We introduce a new
(exponential space) algorithm computing the treewidth of a graph $G$  on
$n$ vertices in time $\cO(\nrPmc^n)$ and a polynomial space
algorithm computing the treewidth in time $\cO(\boundPol^n)$.  We also show that  if the  treewidth of  $G$ is at most $k$, then it can be computed in time
$\cO(({\frac{2n+k+1}{3})^{k+1}\cdot kn^6})$.
This is a refinement of 
the classical result of Arnborg et al.
Running
times of all these  algorithms
strongly depend on possibilities of  fast
 enumeration of specific structures in a graph, namely, potential maximal cliques,
and minimal separators
\cite{BodlaenderFKKT06,BouchitteT01,BouchitteT02,FominKT04,Villanger06}.
The new combinatorial lemma is crucial in obtaining
 new combinatorial
bounds and enumeration algorithms for minimal separators and
potential maximal cliques, which, in turn, provides faster
algorithms for treewidth.

Similar improvements in running times from $\cO({1.8899}^n)$ to $\cO(\nrPmc^n)$ can
be obtained for a number of results in the literature on problems
related to treewidth (we skip  definitions here). For example, by
combining the ideas from \cite{FominKT04} it is
possible to compute   the fill-in  of a graph in
time $\cO(\nrPmc^n)$. Another example are the treelength and the Chordal Sandwich
problem  \cite{Lokshtanov07} which also can be solved in time $\cO(\nrPmc^n)$ by making use of our technique.

The remaining part of the paper is organized as follows. In the
next section we provide definitions and preliminary results. In
Section~\ref{sec:FL}, we prove our main combinatorial tool.
By making use of this tool, in
Section~\ref{sec:CB}, we prove combinatorial bounds on the number
of minimal separators and potential maximal cliques and obtain algorithm enumerating these structures. These results form  the basis for all our algorithms computing
the treewidth of a graph presented in Sections ~\ref{sec:ESTW},  \ref{sec:tw_k}, and \ref{sec:PSTW}.

\section{Preliminaries}\label{sec:prelim}
We denote by $G=(V,E)$ a finite, undirected and simple graph
with $|V|=n$ vertices and $|E|=m$ edges.
For any non-empty subset $W\subseteq V$, the subgraph of $G$ induced
by $W$ is denoted by $G[W]$. We say that a vertex set $S\subseteq V$ is \emph{connected} if
$G[S]$ is connected.

The \emph{neighborhood} of a vertex $v$ is
$N(v)=\{u\in V:~\{u,v\}\in E\}$
and 
for a vertex set $S \subseteq V$ we set
$N(S) = \bigcup_{v \in S} N(v)\sm S$.
A \emph{clique} $C$ of a graph $G$ is a subset of $V$ such that all the
vertices of $C$ are pairwise adjacent.

\medskip
\noindent\textbf{Minimal separators.}
Let $u$ and $v$ be two non adjacent vertices of a graph $G=(V,E)$. A set of
vertices $S \subseteq V$
is an {\em $u,v$-separator} if $u$ and $v$  are in different connected components
of the graph $G[V \sm S]$. A connected component $C$ of $G[V \sm S]$
is a {\em full} component associated to $S$ if $N(C)=S$.
$S$ is a {\em minimal
$u,v$-separator} of $G$ if no proper subset of $S$ is an $u,v$-separator.
We say that $S$ is a {\em minimal separator}
of $G$ if there are two vertices $u$ and $v$ such
that $S$ is a minimal $u,v$-separator. Notice that a minimal separator can be
strictly included in another one. We denote by $\Delta_G$ the set of all
minimal separators of $G$.

We need the following result due to Berry et al. \cite{BerryBC00} (see also  Kloks et al. \cite{KloksK98})

\begin{proposition}[\cite{BerryBC00}]\label{pr:listing_minsep}
There is an algorithm listing all  minimal separators of an input
graph $G$ in $\cO(n^3|\Delta_G|)$ time.
\end{proposition}
The following proposition is an exercise  in \cite{Golumbic80}.
\begin{proposition}[Folklore]\label{pr:full_components}
A set $S$ of vertices of $G$ is a minimal $a,b$-separator if and
only if $a$ and $b$ are in different full components associated to $S$.
In particular, $S$ is a minimal separator if and only if
there are at least two distinct full components associated to $S$.
\end{proposition}


\medskip
\noindent\textbf{Potential maximal cliques.} A graph $H$ is {\em
chordal} (or {\em triangulated}) if every cycle of length at least
four has a chord, i.e. an edge between two non-consecutive
vertices of the cycle. A {\em triangulation} of a graph $G=(V,E)$
is a chordal graph $H = (V, E')$ such that $E \subseteq E'$. $H$
is a {\em minimal triangulation} if for any  set $E''$
with $E \subseteq E'' \subset E'$, the graph $F=(V, E'')$ is not
chordal.

A set of vertices $\Omega \subseteq V$ of a graph $G$ is called a
{\em potential maximal clique} if there is a minimal triangulation
$H$ of $G$ such that $\Omega$ is a maximal clique of $H$. We
denote by $\Pi_G$ the set of all potential maximal cliques of $G$.

The following result on the structure of potential maximal cliques is due to
Bouchitt\'e and Todinca.
\begin{proposition}[\cite{BouchitteT01}]\label{pr:pmc_sep}
Let $K \subseteq V$ be a set of vertices of the graph $G=(V,E)$. Let
$\cC(K) = \{ C_1(K), \ldots, C_p(K)\}$ be the set of the connected
components of $G[V \sm K]$ and let ${\mathcal S}(K) = \{ S_1(K) ,$ $S_2(K) , \ldots  , S_p(K)\}$ where $S_i(K)$, $i\in \{1,2,\ldots ,p\}$,
is the set of those
vertices of $K$ which are adjacent to at least one vertex of the
component $C_i(K)$.
Then $K$ is a potential maximal clique of $G$ if and only if:
\begin{enumerate}
\item $G [V\sm K]$ has no full component associated to $K$, and
\item the graph on the vertex set $K$ obtained from $G[K]$ by
completing each $S_i \in {\mathcal{S}}(K)$
into a clique, is a complete graph.
\end{enumerate}
\end{proposition}
The following result is also due to Bouchitt\'e and Todinca.
\begin{proposition}[\cite{BouchitteT01}]\label{pr:pmc_rec}
There is an algorithm that, given a graph $G = (V,E)$ and a set of
vertices $K \subseteq V$, verifies if $K$ is a potential maximal
clique of $G$. The time complexity of the algorithm
is $\cO(nm)$.
\end{proposition}

\noindent\textbf{Treewidth.}
 A {\em tree decomposition} of a graph $G=(V,E)$ is a pair $(\chi, T)$ in which $T=(V_T, E_T)$ is a tree and
$\chi=\{\chi_i|i\in V_T\}$ is a family of subsets of $V$  such
that: (1) $\bigcup_{i\in V_T}\chi_i= V$; (2) for each edge $e=\{u,
v\} \in E$ there  exists an $i\in V_T$ such that both $u$ and $v$
belong to $\chi_i$; and (3) for all $v\in V$, the set of nodes
$\{i\in V_T|v \in \chi_i\}$ forms a connected subtree of $T$.
To distinguish between vertices of the original graph $G$ and vertices
of $T$, we call vertices of $T$ {\em nodes} and their
corresponding $\chi_i$'s {\em bags}. The maximum size of a bag
 minus one is called the {\em width} of the tree
decomposition.
The {\em treewidth} of a graph $G$, $tw(G)$,  is the
minimum width over all possible tree decompositions of $G$.

An alternative
definition of treewidth is via minimal triangulations.
The  \emph{treewidth} of a graph $G$ is
 the minimum of $\omega(H) - 1$ taken over all triangulations $H$ of $G$.
(By  $\omega (H)$ we denote the  
maximum clique-size of a graph $H$.)

Our algorithm for treewidth is based on the following result.

\begin{proposition}[\cite{FominKT04}]\label{pr:comp_tw_mfi}
There is an algorithm that, given a graph $G$ together with the
list of its minimal separators $\Delta_G$  and the list
of its potential maximal cliques $\Pi_G$, computes the treewidth  of $G$ in
$\cO(n^3 \, (|\Pi_G| + |\Delta_G|)$ time.
Moreover, the algorithm constructs an optimal triangulation for
the treewidth.
\end{proposition}

\section{Combinatorial Lemma}
\label{sec:FL}
The following lemma is our main combinatorial tool.

\begin{lemma}[Main Lemma]\label{le:connectedComp} Let $G=(V,E)$ be a graph. For every $v\in
V$, and $b,f\geq 0$,  the number of connected vertex subsets
$B\subseteq V$ such that
\begin{itemize}
  \item[$(i)$] $v \in B$,
  \item[$(ii)$] $|B| = b+1$, and
  \item[$(iii)$] $|N(B)|=f$
\end{itemize}
 is at most $\binom{b+f}{b}$.
\end{lemma}
\begin{proof}
Let $v$ be a vertex of a graph $G=(V,E)$.
 For $b+f=0$ Lemma trivially holds.
  We proceed by induction assuming that
for some $k > 0$ and every $b$ and $f$ such that $ b+f\leq k-1$, Lemma
holds.    For $b$ and $f$ such that $b+f=k$ we define $\mB$ as the
set of sets $B$ satisfying $(i),(ii),(iii)$. We claim that
\[
|\mB | \leq \binom{b+f}{b}.
\]
Since the claim always holds for $b=0$, let us assume that $b>0$.

Let $N(v)=\{v_1, v_2, \dots, v_p\}$. 
For $1\leq i \leq p$, we define  $\mB_i$ as the set of all connected
subsets $B$ such that
\begin{itemize}
 \item Vertices $v, v_i \in B$,
 \item For every $j<i$, $v_j\not\in B$,
 \item $|B| = b+1$,
 \item  $|N(B)|=f$.
\end{itemize}
Let us note, that every set $B$ satisfying the conditions of the
lemma is in some set $\mB_i$ for some $i$, and that for $i\neq j$,
$\mB_i \cap \mB_j=\emptyset$. Therefore,
\begin{eqnarray}\label{eq:summing_firebreaks}
|\mB|=\sum_{i=1}^{p}{|\mB_i|}.
\end{eqnarray}

For every $i> f+1$, $|\mB_i|=0$ (this is because for every $B\in
B_i$, the set $N(B)$ contains vertices $v_1,\dots, v_{i-1}$ and
thus is of size at least $f+1$.) Thus
(\ref{eq:summing_firebreaks}) can be rewritten  as follows
\begin{eqnarray}\label{eq:eq1}
|\mB|=\sum_{i=1}^{f+1}{|\mB_i|}.
\end{eqnarray}

Let $G_i$ be the graph obtained from $G$  by contracting edge
$\{v,v_i\}$ (removing the loop, reduce double edges to single edges, and calling the new vertex by $v$) and removing vertices $v_1,\dots, v_{i-1}$. Then the
cardinality of $\mB_i$ is equal to the number of the connected
vertex subsets $B$ of $G_i$ such that
\begin{itemize}
 \item $v \in B$,
 \item  $|B| = b$,
 \item  $|N(B)|=f-i+1$.
\end{itemize}
By the induction assumption, this number is at most
$\binom{f+b-i}{b-1}$ and (\ref{eq:eq1}) yields that
\[
|\mB|=\sum_{i=1}^{f+1}{|\mB_i|}\leq  \sum_{i=1}^{f+1}
\binom{f+b-i}{b-1}=\binom{b+f}{b}.
\]
\qed \end{proof}

The inductive proof of the Main Lemma can be easily turned
into a recursive polynomial space enumeration algorithm (we skip the proof here).

\begin{lemma}\label{le:connected_enumerate}
All vertex sets of size $b+1$ with $f$ neighbors in  a graph $G$ can be enumerated in time
  $\cO(n\binom{b+f}{b})$ by making use of  polynomial space.
\end{lemma}

\section{Combinatorial bounds}\label{sec:CB}
In this section we provide combinatorial bounds on the number of
minimal separators and potential maximal cliques in a graph. Both
bounds are obtained by applying the Main Lemma on the respectice problems.

\subsection{Minimal separators}

\begin{theorem}\label{th:sepNr}
Let $\Delta_G$ be the set of all minimal separators in a graph $G$ on $n$ vertices. Then
$|\Delta_G|= \cO(\nrSep^n)$.
\end{theorem}
\begin{proof}
For   $1 \leq i \leq n$, let $f(i)$ be the number of all minimal separators in
$G$ of size $i$.
Then
\begin{eqnarray}\label{eq:int1}
|\Delta_G |=\sum_{1}^{n}f(i).
\end{eqnarray}

Let $S$ be a minimal separator of size  $\alpha n$, where $0 < \alpha < 1$.
By Proposition~\ref{pr:full_components},
there exists two full components $C_1$ and $C_2$ associated to $S$.
Let us assume that $|C_1| \leq |C_2|$.
Then $|C_1| \leq (1-\alpha)n/2$. From the definition of a full component $C_1$
associated to $S$, we have that $N(C_1) = S$. Thus, $f(\alpha n)$ is at most
the number of connected vertex sets $C$ of size at most  $(1-\alpha)n/2$
with neighborhoods of size $|N(C)| = \alpha n$. Hence, to bound
$f(\alpha n)$ we can use the Main Lemma for every vertex of
$G$.

By Lemma~\ref{le:connectedComp}, we have that for every vertex $v$, the number of full components
of size $b+1  = (1-\alpha)n/2 $ containing  $v$ and with  neighborhoods of size
 $\alpha n$ is at most
\[
\binom{b+\alpha n}{b} \leq \binom{(1+\alpha)n/2}{b}.
\]
Therefore
\begin{equation}
f(\alpha n) \leq
n \cdot \sum_{i = 1}^{(1-\alpha)n/2} \binom{i + \alpha n}{i} <
n \cdot \sum_{i = 1}^{(1-\alpha)n/2} \binom{(1+\alpha)n/2}{i}.
\label{eq:boundonminsep}
\end{equation}

For $\alpha \leq 1/3$, we have
\[
\sum_{i = 1}^{(1-\alpha)n/2} \binom{(1+\alpha)n/2}{i} <
2^{(1+\alpha)n/2} <
2^{2n/3} <
1.59^n,\]
and thus
\begin{eqnarray}\label{eq:int2}
\sum_{i=1}^{n/3}f(i)\ = \cO(1.59^n).
\end{eqnarray}

For $\alpha \geq 1/3$, 
one can use the well known fact that the sum
$\sum_{k=1}^{\lfloor j/2\rfloor} \binom{j-k}{k}$ is equal to the 
$(j+1)$-st Fibonacci number to show that 



\[
\sum_{i = 1}^{(1-\alpha)n/2} \binom{(1+\alpha)n/2}{i} <
n \cdot \varphi^n,
\]
where  $\varphi = (1+ \sqrt{5})/2< \nrSep^n$ is the golden ratio.
%
%

Therefore,

\begin{eqnarray}\label{eq:int3}
\sum_{i=n/3}^{n}f(i)\ = \cO(\nrSep^n).
\end{eqnarray}
Finally, the theorem follows from the formulas (\ref{eq:int1}),(\ref{eq:int2}) and (\ref{eq:int3}).
\qed \end{proof}

\subsection{Potential maximal cliques}\label{page:weblink}

\begin{definition}[\cite{BouchitteT01}]\label{de:active_inactive}
Let $\Omega$ be a potential maximal clique of a graph $G$
and let
$S \subset \Omega$ be a minimal separator of $G$. We say that $S$ is an
{\em active separator for $\Omega$}, if $\Omega$ is not a clique
in the graph  obtained
from $G$ by completing all the minimal separators contained in $\Omega$,
except $S$.
A potential maximal clique $\Omega$
containing an active separator (for $\Omega$) is called
a  \emph{nice potential maximal clique}.
\end{definition}

We need the following result by Bouchitt\'e and Todinca.
\begin{proposition}[\cite{BouchitteT02}]\label{pr:nicePMClist}
  \label{type_pmc_theorem}
  Let $\Omega$ be a potential maximal clique of $G=(V,E)$, let $u$ be a vertex of $G$,
  and let $G'=G[V \setminus \{u\}]$.
  Then one of the following holds:
  \begin{enumerate}
  \item{Either $\Omega$, or $\Omega \setminus \{u\}$ is a potential maximal clique of $G'$;}
  \item{$\Omega = S \cup \{u\}$, where $S$ is a minimal separator of $G$;}
  \item{$\Omega$ is a nice potential maximal clique.}
  \end{enumerate}
\end{proposition}

Let $\Pi_n$ be  the maximum number of
nice potential maximal cliques that can be contained in a graph on $n$ vertices.
Proposition~\ref{pr:nicePMClist} is useful to bound the number of potential maximal cliques
in a graph by the number of minimal separators $\Delta_G$ and $\Pi_n$.

\begin{lemma}\label{le:pmc_bound}
For any graph $G=(V,E)$,
$|\Pi_G|\leq n(n|\Delta_G| + \Pi_n)$.
\end{lemma}
\begin{proof}
Let $v_1,v_2,...,v_n$ be an ordering of  $V$ and
let $V_i = \bigcup_{j=1}^i v_j$. The proof of the lemma follows from the following claim
$\Pi_{G[V_{i+1}]} \leq \Pi_{G[V_{i}]} + n|\Delta_G| + \Pi_n$
which can be proved by making inductive use of Proposition~\ref{pr:nicePMClist}.
\qed \end{proof}

\begin{definition}
Let $\Omega\in \Pi_G$,   $v\in \Omega$, and  $C_{v}$ be the connected component of
$G[V \setminus (\Omega \setminus \{v\})]$ containing $v$.
We call the pair $(C_{v},v)$ by
 \emph{vertex representation} of $\Omega$.
\end{definition}

\begin{lemma}\label{le:vertexRep}
Let $(C_{v},v)$ be a vertex representation of $\Omega$.
Then $\Omega = N(C_{v}) \cup \{v\}$.
\end{lemma}

\begin{proof}
By Proposition~\ref{pr:pmc_sep},
every vertex $u \in \Omega \setminus \{v\}$, is
 either adjacent to $v$, or  there exists a connected component $C$ of $G[V \setminus \Omega]$
such that $u,v \in N(C)$.
Since  $C \subset C_{v}$, we have that
$\Omega \setminus \{v\} \subseteq N(C_{v})$.
Every connected component $C$ of $G[V \setminus \Omega]$ that
contains $v \in N(C)$ is contained in $C_{v}$ and
$N(C) \subset \Omega$ for every $C$, therefore
$\Omega \setminus \{v\} = N(C_{v})$.
\qed \end{proof}

We need also the following result from \cite{Villanger06}.
\begin{proposition}[\cite{Villanger06}]\label{pr:vertex_rep_2_3}
Let $\Omega$ be a nice potential maximal clique of size
$\alpha n $ in a graph $G$. There exists a vertex representation
$(C_{v},v)$ of $\Omega$ such that $|C_{v}| \leq \lceil \frac{2(1-\alpha)n}{3} \rceil$.
\end{proposition}

Now everything is settled to apply Main Lemma.
\begin{lemma}\label{le:nice_pmc_bound}
The number of nice potential maximal cliques in a
graph $G=(V,E)$ is  $\cO(\nrPmc^n)$.
\end{lemma}
\begin{proof}
 By Proposition~\ref{pr:vertex_rep_2_3},
for every nice potential maximal clique
 $\Omega$ of cardinality $\alpha n$, there exists a vertex representation $(C_{v},v)$ of $\Omega$
such that $|C_{v}| \leq \lceil 2n(1-\alpha)/3 \rceil$.
Let $b+1$ be the number of vertices in $C_{v}$.
By Lemma~\ref{le:connectedComp}, for every vertex $v$, the number of such pairs
$(C_{v},v)$  is at most
\begin{equation*}
\sum_{i=1}^{{2(1-\alpha)n}/{3}} \binom{{(2+\alpha)n}/{3}}{i}.
\end{equation*}

As in the proof of Theorem~\ref{th:sepNr}, 
for $\alpha \leq 2/5$ 
the above sum is  $\cO(1.7549^n)$. For $\alpha \geq 2/5$, by making use of the fact that 
$\sum_{k=1}^{\lfloor j/2\rfloor} \binom{j-k}{2k}$ is equal to the
$(j+1)$-st number of the sequence $\{a_i\}_{i=0}^{\infty}$ such that $a_{i}= 2 a_{i-1} -a_{i-2} + a_{i-3}$, with $a_0=0$, $a_1=1$, and $a_2=2$, it is possible to show that
the value of the above sum, and thus the number of nice potential maximal cliques, is
 $\cO(1.7549^n)$.
\qed \end{proof}

By combining Lemma~\ref{le:pmc_bound}, \ref{le:nice_pmc_bound}  and Theorem~\ref{th:sepNr}
we arrive at the main result of this subsection.
\begin{theorem} For any graph $G$, $|\Pi_G|=\cO(\nrPmc^n)$.
\end{theorem}

\section{Exponential space exact algorithm for treewidth}\label{sec:ESTW}
Our algorithm computing the  treewidth of a graph is based on Proposition~\ref{pr:comp_tw_mfi}.
By making use of 
Proposition~\ref{pr:comp_tw_mfi} we need to know how to  
list minimal separators and potential maximal cliques.
By Proposition~\ref{pr:listing_minsep} and Theorem~\ref{th:sepNr},  all minimal separators can be listed in
time $\cO(\nrSep^n)$. 
The proof of the following lemma is postponed till the full version of this paper. 

\begin{lemma}\label{le:listing_pmc}
 For any graph $G$ on $n$ vertices,
the set of potential maximal cliques can be listed in $\cO(\nrPmc^n)$ time.
\end{lemma}

As an immediate corollary of  Proposition~\ref{pr:listing_minsep} and
Lemma~\ref{le:listing_pmc}, we have the following result.

\begin{theorem}
The treewidth of a graph on $n$ vertices can be computed in time
 $\cO(\nrPmc^n)$.
\end{theorem}



\section{Computing treewidth at most $k$}\label{sec:tw_k}
In this section we show how the lemma bounding the number of connected components
can be used to refine the  classical result  of
Arnborg et al.  \cite{ArnborgCP87}.

By  Proposition~\ref{pr:comp_tw_mfi},   the treewidth of a graph  can be computed in $\cO(n^3 \, (|\Pi_G| + |\Delta_G|))$ time if
the list of all minimal separators $\Delta_G$  and the list of all
potential maximal cliques $\Pi_G$ of $G$ are  given.
Actually, the results of Proposition~\ref{pr:comp_tw_mfi} can be strengthened  (with almost the same proof  as in \cite{FominKratschTodincaV08})
as follows. Let $\Delta_G[k]$  be the set of minimal separators
and let $\Pi_G[k]$ be the set of  potential maximal cliques of size at most $k$.
 \begin{lemma}\label{lem:refinedcomptw_mfi}
 Given a graph $G$ with sets   $\Delta_G[k]$ and   $\Pi_G[k+1]$,  it can be decided in time  $\cO(n^3 \, (|\Pi_G[k+1]| + |\Delta_G[k]|))$ if the treewidth of $G$ is at most $k$.
 Moreover, if the treewidth of $G$ is at most $k$, an optimal tree decomposition can be constructed within the same time.
 \end{lemma}


%

By  Lemma~\ref{le:connected_enumerate} and Equation~(\ref{eq:boundonminsep}),
\begin{equation}
  |\Delta_G[k]|\leq kn \cdot \sum_{i = 1}^{(n-k)/2} \binom{(n+k)/2}{i} \leq kn^2 \cdot  \binom{(n+k)/2}{k},
  \label{eq:kseps}
  \end{equation}
and it is possible to list all vertex subsets containing all separators from $\Delta_G[k]$ in time
$
\cO( kn^2 \cdot  \binom{(n+k)/2}{k})).
$
 For each such a subset one can check in time $\cO(n^2)$ if it is a minimal separator or not, and thus all minimal separators of size at most $k$ can be
listed in time
$
\cO(kn^4  \cdot  \binom{(n+k)/2}{k}).
$

Let $\Pi_n[k]$ be the maximum number of nice potential maximal cliques of size at most $k$ that can be in a graph on $n$ vertices.
By Proposition~\ref{pr:vertex_rep_2_3},
\[
|\Pi_n[k]|\leq kn \cdot \sum_{i = 1}^{(n-k)2/3} \binom{(2n+k)/3}{i} \leq kn^2 \cdot \binom{(2n+k)/3}{k},
\]
and by making use of Proposition~\ref{pr:pmc_rec},
all
 nice potential maximal cliques of size at most $k$  can be listed in time $ \cO(kn^5 \cdot \binom{(2n+k)/3}{k})$.

Finally, we use nice potential maximal cliques and minimal separators of size $k$ to generate all potential maximal cliques of size at most $k$.

\begin{lemma}\label{le:pmc_bound_k}
For every graph $G$ on $n$ vertices,
$|\Pi_G[k]| \leq n(|\Delta_G[k]| +\Pi_n[k] )$
and
all
potential maximal cliques of  $G$ of size at most $k$ can be listed in time
 $ \cO(kn^6 \cdot \binom{(2n+k)/3}{k})$.
\end{lemma}
\begin{proof}
Let $v_1,v_2,...,v_n$ be an ordering of  $V$ and
let $V_i = \bigcup_{j=1}^i v_j$.
By Proposition~\ref{pr:nicePMClist} and Lemma~\ref{le:pmc_bound}, every
potential maximal clique of $G[V_i]$  either is  a nice potential maximal clique of  $G[V_i]$,  or is a potential maximal clique
of $G[V_{i-1}]$, or is obtained by adding  $v_i$ to a minimal
separator or  a potential maximal clique of $G[V_{i-1}]$.
This yields that $|\Pi_G[k]| \leq n(|\Delta_G[k]| +\Pi_n[k] )$. To   list
all potential maximal cliques, for each $i$, $1\leq i\leq n$, we list all minimal separators and nice potential maximal cliques in $G[V_i]$. This can be done in
 time $ \cO(kn^6 \cdot \binom{(2n+k)/3}{k})$. The total number of all such structures is at most   $ kn^3 \cdot \binom{(2n+k)/3}{k}$.  By making use of dynamic programing,
one can check if  adding $v_i$ to a minimal separator or potential maximal clique of   $G[V_{i-1}]$ creates a potential maximal clique in $G[V_{i}]$, which by
 Proposition~\ref{pr:pmc_rec}   can be done in time $\cO(n^3)$.  Thus, dynamic programming can be done in $ \cO(kn^6 \cdot \binom{(2n+k)/3}{k})$ steps.
\qed \end{proof}

Now putting Lemma~\ref{lem:refinedcomptw_mfi},  Lemma~\ref{le:pmc_bound_k} and  Equation (\ref{eq:kseps}) together, we obtain the main result of this section.
\begin{theorem}
There exists an algorithm that for a given  graph $G$ and integer $k\geq 0$, either
computes a tree decomposition of $G$
of the minimum width,  or correctly concludes that the treewidth of $G$ is at least
$k+1$. The running  time of this algorithm is $\cO(kn^6 \cdot\binom{(2n+k+1)/3}{k+1} )=\cO(kn^6 \cdot  ({\frac{2n+k+1}{3}})^{k+1})$ .
\end{theorem}
\begin{proof}
By the previous discussions in this section we can list all the minimal separators
and potential maximal cliques of size at most $k+1$ in $O^*(\binom{(2n+k)/3}{k})$ time.
These minimal separators and potential maximal cliques are then used as input
to the dynamic programming algorithm of \cite{FominKT04}.
\qed \end{proof}





\section{Polynomial space exact algorithm for treewidth}
\label{sec:PSTW}
The algorithm used in Proposition~\ref{pr:listing_minsep} requires  exponential space because
it is based on dynamic programming which keeps a table with all potential maximal cliques.
As a consequence our $\cO(\nrPmc^n)$ time algorithm for computing the treewidth also uses $\cO(\nrPmc^n)$ space.

When restricting to polynomial space, we cannot
store all the minimal separators and all the potential maximal cliques.
The idea used to avoid this is to search for a
``central'' potential maximal clique or a minimal separator
in the graph which can safely be completed into a clique. A similar
idea is used in \cite{BodlaenderFKKT06}, however the improvement
in the running time of our algorithm, is due to the following
lemma and the technique used for listing minimal separators. Both
results are, again, based on the Main Lemma.

\begin{lemma}\label{le:findPMC}
For a given graph $G=(V,E)$ and $0<\alpha<1$,  one can list  in time
$\cO(mn^2 \cdot 2^{n(1-\alpha)})$ and polynomial space all potential
maximal cliques of $G$ such that for every potential maximal clique
$\Omega$ from this list, there is
a connected component of $G[V \setminus \Omega]$ of size  at least $\alpha n$.
\end{lemma}
\begin{proof} Let $\Omega$ be a potential maximal clique
satisfying the conditions of the lemma, and let $C$ be the connected component of size
at least $\alpha n$.
By Proposition \ref{pr:pmc_sep}, $N(C)$ is a minimal separator contained in $\Omega$ and
$\Omega \setminus N(C) \neq \emptyset$.
Let $(C_u,u)$ be a vertex representation of $\Omega$, where $u \in \Omega \setminus N(C)$.
 Since $u$ is not adjacent to any vertex in $C$, we have that  $C_u \cap C = \emptyset$.
To find $\Omega$, we try to find its vertex representation by
a connected vertex set such that the closed neighborhood of this
set is of size at most $n(1-\alpha)$. By the Main Lemma,
the number of such sets is at most
\[
n \cdot \sum_{i=1}^{n(1-\alpha)} \binom{n(1-\alpha)}{i} = n \cdot 2^{n(1-\alpha)},
\]
and by Lemma \ref{le:connected_enumerate}, all these sets can be listed in
$\cO( n \cdot 2^{n(1-\alpha)})$ steps and within polynomial space. Finally, for each
set we use Lemma~\ref{le:vertexRep} and Proposition~\ref{pr:pmc_rec} to check  in time $\cO(mn)$
if the set is a potential maximal clique.
\qed \end{proof}

We also use the following result which is a slight modification of the result from
\cite{BodlaenderFKKT06}, where it is stated in terms of elimination orderings.

\begin{proposition}[\cite{BodlaenderFKKT06}]\label{pr:pol_spaceTW}
For a given graph $G=(V,E)$ and a clique $K \subset V$,
there exists a polynomial space algorithm, that computes the
optimum tree decomposition $(\chi,T)$
of $G$, subject to the condition
that the vertices of $K$ form a bag which is a leaf of $T$.
This algorithm runs  in time $\cO^*(4^{n-|K|})$.
\end{proposition}

\begin{theorem}\label{thm:polspace}
The treewidth of a graph $G=(V,E)$ can be computed in
$\cO(\boundPol^n)$ time and polynomial space.
\end{theorem}
\begin{proof}
It is well known (and follows from the properties of clique trees of chordal graphs),
that there is an optimal tree decomposition $(\chi,T)$,
$\{\chi_i \colon i \in V_T \}, T=(V_T,E_T)$, of $G$, where every bag is a
potential maximal clique \cite{BouchitteT01,Buneman74,HoL89}. Among all the bags of
 $\chi$, let $\chi_i$ be a bag such that the largest connected component of
$G[V \setminus \chi_i]$ is of minimum size, i.e. $\chi_i$ is a bag with the minimum value of
\[
\max\{ |C| \colon C \text{ is a connected component of } G[V \setminus \chi_i]\},
\]
where  minimum is
taken over all bags of $\chi$. Let $C_i$ be the connected component of $G-\chi_i$ of maximum size.

Our further strategy depends on the size of $|C_i|$.
Let us assume first that $|C_i| < \const n$. In this case, by  Lemma \ref{le:findPMC},
the set  of potential maximal cliques $\cal{S}$ such that for every $\Omega\in \cal{S}$
the maximum size of  a component of $G[V \setminus \Omega]$  is
$|C_i|$,  can be listed in time
$\cO(mn^2 \cdot 2^{n-|C_i|})$   and polynomial space.
 Since $\chi_i\in \cal{S} $,  we have that there is a potential maximal clique
 $\Omega \in \cal{S} $ such that  $tw(G_{\Omega})=tw(G)$, where   $G_{\Omega}$ is obtained
 from $G$ by turning $\Omega$ into a clique. The treewidth of $G_{\Omega}$ is equal to the
maximum of minimum width of decompositions of $G_{\Omega}[C\cup \Omega]$ with $\Omega$ forming
a leaf bag, where $C$ is a
connected component of $G_{\Omega}[V \setminus \Omega]$. Let us remind that the size of each such
component is at most $|C_i|$.

By Proposition~\ref{pr:pol_spaceTW}, the optimum width of $G_{\Omega}[C\cup \Omega]$
for every connected component $C$ of $G_{\Omega}[C\cup \Omega]$
 (and with  $\Omega$ forming
a leaf bag)  can be computed in
$\cO^*(4^{|C|})=\cO^*(4^{|C_i|})$, time and thus
the treewidth of $G$ can be found in time
\[
\cO^*(2^{n-|C_i|} \cdot 4^{|C_i|}) =
\cO^*(2^{(1-\const)n} \cdot 4^{\const n}) = \cO(\boundPol^n).
\]
Thus if  $|C_i| < \const n$,  we  compute the treewidth of $G$, and the running time of
this polynomial space procedure is  $\cO(\boundPol^n)$.

\medskip

 Let us consider the case $|C_i| \geq \const n$.
For each connected component $C$ of $G[V \setminus \chi_i]$, there exists a bag $\chi_{i'} \subset N(C) \cup C$ and
a minimal separator $S = \chi_i \cap \chi_{i'}$ in $\chi_i$ that separates $C$ from the rest of
the graph. Let $S = \chi_i \cap \chi_j$ be the separator in $\chi_i$ that separates
$C_i$ from the rest of the graph.
Let $G_S$ be the graph obtained from $G$ by turning $S$ into a clique. Then $tw(G_S)=tw(G)$.
To compute the  treewidth of  $G_S$ we
compute the
minimum width of decompositions of $G_{S}[C\cup S]$ with $S$ forming
a leaf bag, where $C$ is a
connected component of $G_{S}[V \setminus S]$, and then take the maximum of these values.

By the definition of $\chi_i$, there exists a connected component $C_j$ of
$G[V \setminus \chi_j]$, such that $|C_j| \geq |C_i|$.
By Proposition~\ref{pr:pmc_sep}, $\chi_j \not\subseteq \chi_i $.
Thus  $\chi_j \setminus \chi_i \neq \emptyset$, and the size of every connected
component in $G[C_i \setminus \chi_j]$ is at most  $|C_i|-1$.
Furthermore,  since $S = \chi_i \cap \chi_j$, we have that
every connected component of $G[C_i \setminus \chi_j]$ is
also a connected component of $G[V \setminus \chi_j]$. This yields that
$C_j \cap C_i = \emptyset$ and
 that both $C_i$ and $C_j$ are full connected components assosiated to $S$.
 Thus $|C_j| + |C_i| \leq n-|S|$.
Every connected component of $G[V \setminus S]$, except   $C_i$, is a
connected component of $G[V \setminus \chi_j]$. Because
 $|C_i| \leq |C_j|$, this implies that
$C_j$ is the largest component of $G[V\setminus S]$. Both $C_i$
and $C_j$ contain at least $\const n$ vertices, thus the size of
$S$ is at most $n(1-2\cdot\const)=\constB n$. By the algorithmic
version of Main Lemma, all sets of such size (and which
form the neighborhood of a set of size $|C_i|$) can be listed in
polynomial space and time
\[
  \cO(nm \cdot \sum_{p = 1}^{\constB n}\binom{|C_i|+p}{p} ).
\]
By Proposition~\ref{pr:pol_spaceTW}, we can
compute the
minimum width of decompositions of $G_{S}[C\cup S]$ with $S$ forming
a leaf bag, where $C$ is a
connected component of $G_{S}[V \setminus S]$, in time
 \[
 \cO^*(4^{|C|})=\cO^*(4^{|C_j|})
 \]
 and polynomial space. Because
 $|C_j|  \leq n-|S|-  |C_i|$, we have that for $|S|=p$,
\[
\cO^*(4^{|C_j|})=\cO^*(4^{n-|C_i|-p}).
 \]

Thus to compute the treewidth of $G_S$ (and the treewidth of $G$), we list all sets
$S$ and for each such a set we use Proposition~\ref{pr:pol_spaceTW} for all graphs
$G_{S}[C\cup S]$.
The running time of this procedure is
\[
 \cO^*(\sum_{p = 1}^{\constB  n}\binom{|C_i|+p}{p}
 \cdot 4^{n - |C_i|-p}).
\]

By Vandermonde's identity, we have that
\[
 \binom{|C_i|+p}{p}=\sum_{k=0}^{p}\binom{\const n +p}{k}\binom{|C_i|-\const n}{k} <
 \sum_{k=0}^{p}\binom{\const n +p}{k} 2^{|C_i|-\const n}.
\]
Thus
\begin{eqnarray*}
  \sum_{p = 1}^{\constB  n}\binom{|C_i|+p}{p}
 \cdot 4^{n - |C_i|-p}  & <&  \sum_{p = 1}^{\constB  n}  \sum_{k=0}^{p}\binom{\const n +p}{k} 2^{|C_i|-\const n}  \cdot 4^{n - |C_i|-p}\\
& \leq  & \sum_{p = 1}^{\constB n} p \binom{\const n+p}{p} \cdot 2^{2((1-\const)n -p)} = \cO(\boundPol^n)
\end{eqnarray*}

%

%
%
%
%
To conclude, if  $|C_i| \geq\const n$,  we  compute the treewidth of $G$ in polynomial space within
 $\cO(\boundPol^n)$ steps.
\qed \end{proof}

\medskip\noindent{\textbf{Acknowledgement.}}
We are grateful to Saket Saurabh for many useful comments,
and to the anonymous referee pointing out that one of the
bounds matched the golden ratio.

%

\begin{thebibliography}{10}

\bibitem{ArnborgCP87}
{\sc S.~Arnborg, D.~G. Corneil, and A.~Proskurowski}, {\em Complexity of
  finding embeddings in a {$k$}-tree}, SIAM J. Algebraic Discrete Methods, 8
  (1987), pp.~277--284.

\bibitem{BerryBC00}
{\sc A.~Berry, J.~P. Bordat, and O.~Cogis}, {\em Generating all the minimal
  separators of a graph.}, Int. J. Found. Comput. Sci., 11 (2000),
  pp.~397--403.






\bibitem{Bodlaender96}
{\sc H.~L. Bodlaender}, {\em A linear-time algorithm for finding
  tree-decompositions of small treewidth.}, SIAM J. Comput., 25 (1996),
  pp.~1305--1317.

\bibitem{Bodlaender98}
{\sc H.~L. Bodlaender}, {\em A partial {$k$}-arboretum of graphs with bounded
  treewidth}, Theoretical Computer Science, 209 (1998), pp.~1--45.

\bibitem{BodlaenderFKKT06}
{\sc H.~L. Bodlaender, F.~V. Fomin, A.~M. C.~A. Koster, D.~Kratsch, and D.~M.
  Thilikos}, {\em On exact algorithms for treewidth.}, in ESA, vol.~4168 of
  LNCS, Springer, 2006, pp.~672--683.

\bibitem{Bodlaender08}
{\sc H.~L.~Bodlaender and A.~M.~C.~A.~Koster}, {\em Combinatorial Optimization on Graphs of Bounded Treewidth}, The Computer Journal, to appear.


\bibitem{Bollobas65}
{\sc B.~Bollob{\'a}s}, {\em On generalized graphs}, Acta Math. Acad. Sci.
  Hungar.,  (1965), pp.~447--452.

\bibitem{BouchitteT01}
{\sc V.~Bouchitt{\'e} and I.~Todinca}, {\em Treewidth and minimum fill-in:
  Grouping the minimal separators}, SIAM J. Comput., 31 (2001), pp.~212--232.

\bibitem{BouchitteT02}
\leavevmode\vrule height 2pt depth -1.6pt width 23pt, {\em Listing all
  potential maximal cliques of a graph.}, Theor. Comput. Sci., 276 (2002),
  pp.~17--32.

\bibitem{Buneman74}
{\sc P.~Buneman}, {\em A characterization of rigid circuit graphs}, Discrete
  Math., 9 (1974), pp.~205--212.

\bibitem{DavisLL62}
{\sc M.~Davis, G.~Logemann, and D.~Loveland}, {\em A machine program for
  theorem-proving}, Comm. ACM, 5 (1962), pp.~394--397.

\bibitem{DavisP60-A}
{\sc M.~Davis and H.~Putnam}, {\em A computing procedure for quantification
  theory}, J. Assoc. Comput. Mach., 7 (1960), pp.~201--215.

\bibitem{FeigeHL05}
{\sc U.~Feige, M.~T. Hajiaghayi, and J.~R. Lee}, {\em Improved approximation
  algorithms for minimum-weight vertex separators.}, in STOC, ACM press, 2005,
  pp.~563--572.

\bibitem{FominGrandoniKratsch05beatcs}
{\sc F.~Fomin, F.~Grandoni, and D.~Kratsch}, {\em Some new techniques in design
  and analysis of exact (exponential) algorithms}, Bulletin of the European
  Association for Theoretical Computer Science, 87 (2005), pp.~47--77.

\bibitem{FominKT04}
{\sc F.~V. Fomin, D.~Kratsch, and I.~Todinca}, {\em Exact (exponential)
  algorithms for treewidth and minimum fill-in.}, in ICALP, vol.~3142 of LNCS,
  Springer, 2004, pp.~568--580.

\bibitem{FominKratschTodincaV08}
{\sc F.~V. Fomin, D.~Kratsch, I.~Todinca, and Y.~Villanger}, {\em Exact
  algorithms for treewidth and minimum fill-in}, SIAM J. Comput.,  (accepted).



\bibitem{Golumbic80}
{\sc M.~C. Golumbic}, {\em Algorithmic Graph Theory and Perfect Graphs},
  Academic Press, New York, 1980.

\bibitem{HeldKarp62jsiam}
{\sc M.~Held and R.~M. Karp}, {\em A dynamic programming approach to sequencing
  problems}, Journal of {SIAM}, 10 (1962), pp.~196--210.

\bibitem{HoL89}
{\sc C.-W. Ho and R.~C.~T. Lee}, {\em Counting clique trees and computing
  perfect elimination schemes in parallel}, Inform. Process. Lett., 31 (1989),
  pp.~61--68.

\bibitem{Iwama04beatcs}
{\sc K.~Iwama}, {\em Worst-case upper bounds for {k-SAT}}, Bulletin of the
  European Association for Theoretical Computer Science, 82 (2004), pp.~61--71.

\bibitem{Jukna01Ex}
{\sc S.~Jukna}, {\em Extremal combinatorics with applications in computer
  science}, Springer-Verlag, Berlin, 2001.

\bibitem{KloksK98}
{\sc T.~Kloks and D.~Kratsch}, {\em Listing all minimal separators of a
  graph.}, SIAM J. Comput., 27 (1998), pp.~605--613.

\bibitem{Lokshtanov07}
{\sc D.~Lokshtanov}, {\em On the complexity of computing treelength}, in MFCS, vol.~4708
  of LNCS, Springer, 2007, pp.~276--287.

\bibitem{RobertsonS86}
{\sc N.~Robertson and P.~D. Seymour}, {\em Graph minors. {II}. {A}lgorithmic
  aspects of tree-width}, Journal of Algorithms, 7 (1986), pp.~309--322.



\bibitem{Schoening05stacs}
{\sc U.~Sch\"oning}, {\em Algorithmics in exponential time}, in STACS, vol.~3404 of LNCS, Springer, 2005, pp.~36--43.




\bibitem{TarjanTrojanowski77}
{\sc R.~E. Tarjan and A.~E. Trojanowski}, {\em Finding a maximum independent
  set}, SIAM Journal on Computing, 6 (1977), pp.~537--546.

\bibitem{Villanger06}
{\sc Y.~Villanger}, {\em Improved exponential-time algorithms for treewidth and
  minimum fill-in.}, in LATIN, vol.~3887 of LNCS, Springer, 2006, pp.~800--811.

\bibitem{Woeginger03}
{\sc G.~Woeginger}, {\em Exact algorithms for {NP}-hard problems: A survey}, in
  Combinatorial Optimization - Eureka, you shrink!, vol.~2570 of LNCS,
  Springer-Verlag, Berlin, 2003, pp.~185--207.

\end{thebibliography}
%


\end{document}